\RequirePackage{fix-cm}

\documentclass[smallextended]{svjour3}

\smartqed
\usepackage{graphicx}
\usepackage{epstopdf}

\begin{document}

\title{A Systematic Analysis of the XMM-Newton Background: I. Dataset and Extraction Procedures}
\titlerunning{XMM-Newton Background: Dataset and Extraction Procedures}

\author{Martino Marelli$^1$ \and David Salvetti$^1$ \and Fabio Gastaldello$^1$ \and Simona Ghizzardi$^1$ \and Silvano Molendi$^1$ \and Andrea De Luca$^{1,3,4}$ \and Alberto Moretti$^2$ \and Mariachiara Rossetti$^1$ \and Andrea Tiengo$^{1,3,4}$}
\institute{Martino Marelli
           \at $^{1}$ INAF-IASF Milano, via E. Bassini 15, I-20133 Milano, Italy\\
           \email{marelli@iasf-milano.inaf.it}         
           \and
$^{2}$ INAF-Osservatorio Astronomico di Brera, via Brera 28, I-20121 Milano, Italy
           \and\\
$^{3}$ Istituto Universitario di Studi Superiori, piazza della Vittoria 15, I-27100 Pavia, Italy            
           \and\\
$^{4}$ Istituto Nazionale di Fisica Nucleare, Sezione di Pavia, via A. Bassi 6, I-27100 Pavia, Italy
}

\date{Received: date / Accepted: date}

\maketitle

\begin{abstract}

{\it XMM-Newton} is the direct precursor of the future ESA {\it ATHENA} mission. A study of its particle-induced background provides therefore significant insight for the {\it ATHENA} mission design.
We make use of $\sim$12 years of data, products from the third {\it XMM-Newton} catalog as well as FP7 EXTraS project to avoid celestial sources contamination and to disentangle the different components of the {\it XMM-Newton} particle-induced background.
Within the ESA R\&D AREMBES collaboration, we built new analysis pipelines to study the different components of this background: this covers time behavior as well as spectral and spatial characteristics.
\end{abstract}


\section{Introduction}
\label{intro}

{\it ATHENA}\footnote{http://www.the-athena-x-ray-observatory.eu} (Advanced Telescope for high-ENergy Astrophysics) is the future X-ray mission of the European Space Agency, under development for launch around 2028 \cite{nan13}.
It is the second L2 large class mission within the ESA Cosmic Vision Program.
The direct predecessor of {\it ATHENA} within the ESA science programme is the European Space Agency's X-ray Multi-Mirror satellite {\it XMM-Newton} \cite{jan01}, the second cornerstone of ESA's Horizon 2000 programme, launched on the 10th December 1999 into a highly elliptical orbit. Operating in the same energy band of {\it XMM-Newton}, {\it ATHENA} will face common instrumental and environmental effects.
{\it XMM-Newton} is therefore an exceptional test for the expected effects that will affect the ATHENA mission.

The {\it XMM-Newton} spacecraft is carrying a set of three X-ray CCD cameras, comprising the European Photon Imaging Camera (EPIC), operating in the energy range from 0.2 to 12 keV.
Two of the cameras are Metal Oxide Semi-conductor (MOS) CCD arrays \cite{tur01}. The third X-ray instrument uses pn CCDs and is referred to as the PN camera \cite{str01}. The two types of EPIC cameras differ for the geometry of the CCD arrays and the instrument design and for other properties, like e.g., their readout times. Each CCD is nearly indipendent from the others, allowing for different configurations and can be indipendently shut off, thus resulting in CCD-dependent Good Time Intervals (lists of the time periods in which each CCD is operating correctly).

The EPIC background can be divided into two main components: a cosmic X-ray background (CXB, both of galactic and extragalactic origin) and an instrumental background \cite{gia62,fra14}. The latter component may be further divided into a detector noise component, which becomes important at low energies ($<$ 200 eV) and a second component which is due to the interaction of particles with the structure surrounding the detectors and the detectors themselves. This component is particularly important at high energies (above a few keV). The particle-induced background can be divided into two components: an external $'$flaring$'$ component, characterized by strong and rapid variability, and a second more stable component. The flaring component is currently attributed to soft protons (with energies smaller than a few 100 keV), which are funneled towards the detectors by the X-ray mirrors. The stable component is due to the interaction of high energy particles (with energies larger than some 100 MeV) with the structure surrounding the detectors and possibly the detectors themselves. In this work, we will concentrate on the two latter components: soft proton background (SP) and high-energy-particles induced background (HEPI). Our comprehension of these processes on board {\it XMM-Newton} is still incomplete, with analysis in literature covering from few Ms of data \cite{lec08,car07} up to 44 Ms of data \cite{kun08}. An accurate analysis of a larger data set will lead to improvements on our knowledge of the known components as well as discovering of new components.

The aim of this work is to exploit the entire {\it XMM-Newton} public archive to produce the most complete and clean data set ever used to characterize {\it XMM-Newton} particle-induced background.
In order to do that, we make a conservative energy selection and  data set selection to minimize other contaminants (celestial sources, cosmic X-ray background and instrumental noise). This is described in section \ref{dataset}.
In section \ref{input} we define and disentangle the quiescent and flaring components of the particle-induced background by studying different regions of the detector (inside and outside the Field of View). A region filter is applied in each observation to further reduce the contamination by celestial sources.
Sections \ref{lc}, \ref{spectra} and \ref{image} describe the final products: we compute clean light curves, spectra and images.

This work is part of the AREMBES project (ATHENA Radiation Environment Models and X-Ray Background Effects Simulators\footnote{http://space-env.esa.int/index.php/news-reader/items/AREMBES.html}), aimed at characterizing the effects of focused and non-focused particles on ATHENA detectors both in terms of contributions to their instrumental background and as source of radiation damage.
Several other results of this project are reported in these proceedings \cite{sal16,ghi16,gas16}. \cite{sal16} uses the data presented in this article to characterize the focused part of the {\it XMM-Newton} background, while \cite{ghi16} employs information provided by this work to study in detail the behaviour of the soft-proton-induced background as a function of the position in the terrestrial magnetosphere. In the end, \cite{gas16} focusses on the study and characterization of the behaviour of the high-energy-particles induced background.
The work described in this paper has been performed through newly-developed python scripts. We also made use of HEAsoft tools v.6.19\footnote{https://heasarc.gsfc.nasa.gov/lheasoft/}, the {\it XMM-Newton} Science Analysis Software (SAS) v.14.0\footnote{http://www.cosmos.esa.int/web/xmm-newton/sas} and {\it XMM-Newton} calibration files available at 2016.
As input, we also took part of the intermediate products of European FP7 EXTraS project (Exploring the X-ray Transient and variable Sky\footnote{http://www.extras-fp7.eu/index.php}, \cite{del15}). 

\section{{\bf Selection of data sets and event filters}}
\label{dataset}

Among the three EPIC cameras, MOS cameras are the best-suited to extract the HEPI due to their large out-field-of-view region (see Figure 16 from {\it XMM-Newton} Users Handbook\footnote{http://xmm-tools.cosmos.esa.int/external/xmm\_user\_support/documentation/uhb}). Unexposed area in MOS cameras are typically $\sim$30\% of exposed areas ($\sim$200 arcmin$^2$) while for PN camera out field of view reduces to $\sim$9\% of exposed area ($\sim$60 arcmin$^2$). Also, the PN camera is much more affected by Out-of-Time events (see {\it XMM-Newton} Users Handbook for more details), thus contaminating the unexposed area with photons coming from the exposed area. More important, PN background is not yet clearly characterized with the same details as MOS (nor the different components studied). We therefore exclude PN camera from our analysis.
On March 2005, an event was registered in the EPIC MOS1 instrument, which was attributed to micrometeoroid impacts scattering debris into the focal plane. In the period immediately following the light flash it became apparent that MOS1 CCD6 was no longer recording events. In order to obtain a data set as uniform as possible with time, we exclude MOS1 camera from our analysis. Althought data from PN and/or MOS1 could be in principle extracted and analyzed, we choose to exploit only the stable and conservative MOS2 data set. All the following results are therefore extracted from the EPIC MOS2 camera only.
Future analysis of PN and MOS1 cameras could lead to an even higher statistics, under the hypothesis of a correct treatement of out-of-time events and analysis of systematics of each CCD, respectively.

\begin{figure*}
  \includegraphics[width=1.0\textwidth]{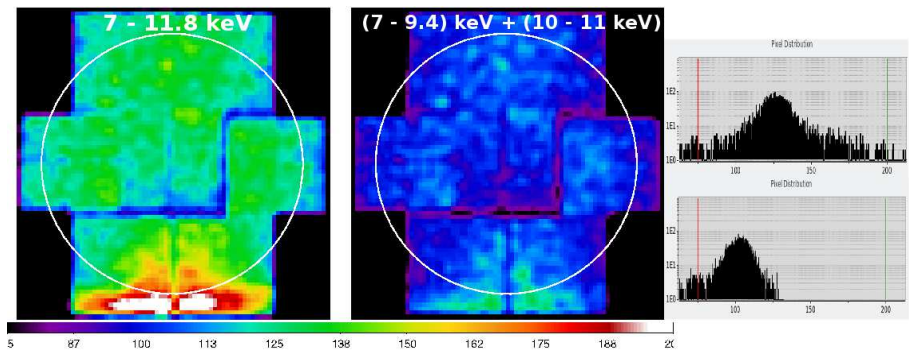}
\caption{Here we show the sum of the images of closed regions in the total 7-11.8 keV band (left) and the selected (7-9.4) + (10-11) keV band (right), respectively. A raw selection of the exposed area is shown with a blue circle. On the right side, we report the distribution of the pixel values for the total band (upper) and selected band (lower) figures. The small shift in the maximum is due to the band sub-selection. The most prominent effect is the absence of the high-counts tail in our selected band due to the counts in the gold line.}
\label{fig-gl}
\end{figure*}

We apply a standard filter on event patterns, using only single and double events. We also make use of the standard flags to avoid bright columns and pixels (FLAG\&0x766a0l63==0, see {\it XMM-Newton} Users Handbook for more details).
In order to minimize the cosmic X-ray background contribution, we make a conservative event selection based on energy. As apparent in Figure B1 of \cite{lec08}, the CXB contribution becomes negligible above 7 keV: we therefore exclude the 0.2-7 keV energy band from our analysis. We also exclude the 11-12 keV energy band due to a prominent instrumental fluorescence line.
In the 7-11 keV energy band, one of the most apparent characteristic of the background spectrum is the gold fluorescence line at 9.7 keV. An analysis of closed observations reveals that such an emission is not spatially uniform, with an excess in CCDs 2 and 7. Through the exclusion of the 9.4-10 energy band, we minimize this effect (see Figure \ref{fig-gl}).

In order to evaluate the contamination from celestial sources, we rely on the 3XMM source catalog distribution 4\footnote{http://xmmssc-www.star.le.ac.uk/Catalogue/3XMM-DR4/}, that analyzes 7598 public {\it XMM-Newton} EPIC exposures made between 2000 February 3 and 2012 December 8. We make a sub-selection of this data set to lower the noise coming from e.g. too bright point sources or extended sources to have a uniform data set, so that results from each observation can be compared.
We make the following selections:\\
a. We make use of intermediate products of the EXTraS project, therefore reducing our data set to the 7190 exposures analyzed at 2016 March.\\
b. In order to avoid problems with the SAS attitude computation (e.g. for the exposure maps), we use only exposures with an attitude stability better than 5$''$, as reported in attitude files.\\
c. To obtain an uniform data set, we select only exposures in the Full Window mode.\\
d. In order to reduce contamination by celestial sources, we rely on the counts flux reported in the 3XMM catalog. We use their hardest band, 4.5-12 keV; under the hypothesis of a power-law spectral model with photon index 2, this flux is reduced to $\sim$ 40\% in our energy band ($\sim$30\% and $\sim$45\% for photon indexes 1 and 3, respectively). We exclude exposures in which the sum of the 3XMM counts flux ($M2\_RATE\_5$) coming from extended sources (source extension $EP\_EXTENT>$12") is higher than 0.05 c s$^{-1}$ (thus $\sim$0.02 c s$^{-1}$ in our band).\\
e. In order to reduce the PSF wings contribution \cite{rea11}, we exclude exposures containing point sources with a mean count flux ($M2\_RATE\_5$) higher than 0.5 c s$^{-1}$.\\
f. In order to avoid problems with the 3XMM source detection due to azimuthal variations of the {\it XMM-Newton} PSF \cite{rea11}, we exclude exposures containing sources with a 0.2-12 keV mean count flux ($M2\_RATE\_8$) higher than 1 c s$^{-1}$.\\
g. In order to exclude the Galactic Center diffuse contribution, we exclude observations centered in the box $\mid b \mid < 20^{\circ}$ and $\mid l \mid < 10^{\circ}$ \cite{kri07}.\\
We call “data set raw” the list of the exposures with filters a, b, c ($\sim$ 143 Ms on 5321 exposures) and “data set clean” the most conservative one, with all the listed filters. Our final data set contains $\sim$ 106 Ms of data on 4342 exposures.

We download Processing Pipeline Subsystem (PPS) {\it XMM-Newton} data sets from the {\it XMM-Newton} Science Archive (XSA) as at the beginning of the AREMBES project (2016, March). These data are automatically processed from observation data files (ODF) using the SAS v 13.5, as reported in the {\it XMM-Newton} Users Handbook.
All the sources definitions, positions, and characteristics are taken from the 3XMM-DR4 catalog.
As input, we also take exposure maps and regions from primary and secondary products of the EXTraS project, as at the beginning of the AREMBES project (2016, March). We note that the regions are optimized to maximize the background contribution and to exclude point-like sources contribution in the 0.2-12 keV energy band. Simulations performed within the framework of the EXTraS project evaluate a residual source contribution $<<$0.5\% of the background contribution in the chosen AREMBES energy band.

\section{{\bf General methods and definitions}}
\label{input}

Our aim is to extract a clean data set and disentagle the different components of the {\it XMM-Newton} particle background. As shown in section \ref{dataset}, the Cosmic X-ray Background and detector noise components can be minimized through an accurate energy, pattern and data set selection. We want to study the two remaining components, soft protons (SP) and and induced by high-energetic particles (HEPI), respectively. The first one is focused by the optics, the second one is not. We therefore extract HEPI events from the detector area that is not exposed to focused particles (out field of view). Analysis of source-free events in the detector area that is exposed to focused particles (in field of view) gives us the informations on SP, after the evaluation of expected HEPI component in this area.

We define as $''$In-Field-of-View$''$ (inFOV) the detector area that is exposed to focused X-ray photons. For the MOS cameras, this area is roughly a 14.5$'$-radius circle composed by seven different squared CCDs, separated by gaps (see Figure 16 from {\it XMM-Newton} User Handbook). In our data set, this selection is obtained by imposing the standard filter flag (FLAG\&0x76ba000)==0, following prescriptions from the 3XMM catalog.
In each observation, we exclude circles around contaminant celestial sources. Circles' radii are taken from (and described within) the EXTraS project.
All the inFOV products are normalized to the total inFOV area, so that the results from different observations are directly comparable.
For each observation, the normalization is based on the integral of exposure map in the excluded regions with respect to the total integral of exposure map, and thus relies on the fundamental hypothesis of a spatial-independent background. Calibration analysis of proton flares already showed a marginally spatial-dependent distribution, peaked around the boresight. To first order, the instrumental background is instead constant throughout the detector \cite{kun08} (for more details see \cite{sal16}).

\begin{figure*}
  \includegraphics[width=0.75\textwidth]{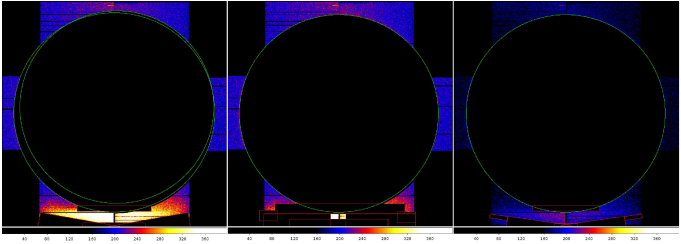}
\caption{This figure shows the different approach that are used for the outFOV area selection. We summed images from our entire data set raw. {\it Left}: this comes from the area and energy selection used by \cite{kun08}. {\it Center}: this comes from the area and energy selection used by \cite{car07}. {\it Right}: this is the outFOV area and energy selection we used for AREMBES.}
\label{fig-of}
\end{figure*}

We define as $''$Out-Field-of-View$''$ (outFOV) the detector area that is not exposed to celestial photons.
For the MOS cameras, this is roughly the total detector area with the exclusion of the inFOV area. In literature, different areas have been used for different studies. \cite{kun08,car07}\footnote{http://xmm2.esac.esa.int/external/xmm\_sw\_cal/background/epic\_scripts.shtml} use an empirical approach based on a limited data sample. \cite{kun08} maximizes the considered outFOV area, so that their results are heavily affected by the Gold Line problem (see Section \ref{dataset}).  \cite{car07} accurately excludes the Gold-Line-affected area, thus greatly reducing the considered outFOV area.
Our long data set allows for an excellent characterization of the areas affected by celestial photons, both in the inFOV and the areas exposed to focused photons due to the holes for the internal calibration source (that are apparent in Figure \ref{fig-of}).
Through an accurate analysis, we excluded such regions from our outFOV.
The resulting region expression is:\\
!((DETX,DETY) IN circle(-50,-180,17540)) \&\&!((DETX,DETY) IN BOX(0,-17000,5900,500,0)) \&\&!((DETX,DETY) IN BOX(0,-20200,2000,500,0)) \&\&!((DETX,DETY) IN BOX(-4800,-20150,5650,915,352)) \&\&!((DETX,DETY) IN BOX(4800,-20150,5650,915,8)) \&\&!((DETX,DETY) IN BOX(-11850,-18600,1575,350,352)) \&\&!((DETX,DETY) IN BOX(11850,-18600,1575,350,8))\\

In order to analyze the HEPI background, we can directly rely on the outFOV data sets. This information allows also for an evaluation of the HEPI background expected in the inFOV area, thus allowing for the analysis of the SP background in the inFOV data sets, after an accurate area normalization.
Thus, when compared to inFOV, outFOV results are normalized to the total inFOV area.
Similar procedures are applied to spectra and images, as described in sections \ref{spectra} and \ref{image} respectively.

\section{Light curves computation}
\label{lc}
As a first step, we evaluate the outFOV to inFOV area rescale factor. The detector active area can be time-dependent due to the instrument degradation. We therefore evaluate the inFOV and outFOV areas during time chunks with at least 3 megaseconds of exposure (chosen to provide a good statistics) thus dividing our data set raw into 47 chunks. For this analysis we use the total 0.2-12 keV energy band.
We sum the counts images of each exposure in a chunk: we define the sum of non-null pixels in the inFOV(outFOV) area as the inFOV(outFOV) area. The excellent statistics provided by 3 Ms of data makes it almost impossible to have zero counts on an active CCD. The time-dependent rescale factor is defined as the ratio of the areas $R_O = outFOV/inFOV$.
The resulting area time variation is not statistically significant, with a mean inFOV area of 674.58$\pm$0.17 arcmins$^2$ and outFOV area of 206.81$\pm$0.17 arcmins$^2$.
We tested the results of this method using the {\tt backscale} SAS tool for a limited subsample of observations. This method works well for simple region files but it can give wrong results for complex regions, and thus we chose to use our more reliable method. In the test cases, we found percentage discrepancies on areas values $<$0.001. Tests on the closed observations data set also revealed a good agreement between the two methods.

We filter each event files for energy, pattern, flag and area selections, as presented in section \ref{input}. For the inFOV region we extract counts only within the EXTraS background regions, thus excluding celestial point sources. Counts from different CCDs are stored separately. For each CCD we produce filtered raw light curves with a set of time bins (10s-500s-5000s).
We extract the Good Time Intervals (GTI) of each CCD from the event file. For each time bin of the raw light curves we compute its GTI fractional coverage (FRACEXP), where 0 means no coverage and 1 full coverage.
In order to correct for the excluded inFOV areas, we rely on EXTraS inFOV exposure maps (they can be computed only for the inFOV area using SAS). They are not corrected for photons vignetting, as needed for our photons+particle-induced background.
We compute the integral of the EXTraS exposure map and the integral of the cheesed exposure map, using the EXTraS background region. The ratio of the two integrals is defined as $R_I$ factor, that can be used to rescale the inFOV counts and therefore $''$fill the holes$''$ due to point-like sources in the inFOV.

\begin{figure*}
  \includegraphics[width=0.75\textwidth]{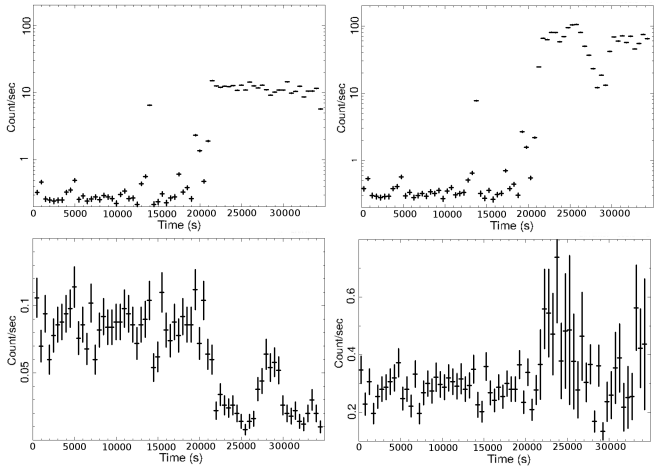}
  \caption{Here we show examples of raw and clean light curves for the MOS2 exposure 0506130201-S002. In this particular case, the presence of numerous CCD-dependent Bad Time Intervals makes our CCD-dependent, time-resolved analysis really effective both in the inFOV and outFOV regions. {\it Upper left}: raw inFOV light curve, as obtained using the evselect SAS command. {\it Upper right}: clean AREMBES inFOV light curve, directly comparable with curves from all the other exposures. {\it Lower left}: raw outFOV light curve, as obtained using the evselect SAS command. {\it Lower right}: clean AREMBES rescaled outFOV light curve; the re-normalization factor makes this curve directly comparable with the inFOV clean light curve. \cite{sal16,ghi16,gas16} will analyse these data in details.}
\label{fig-lc}
\end{figure*}

\begin{figure*}
  \includegraphics[width=0.75\textwidth]{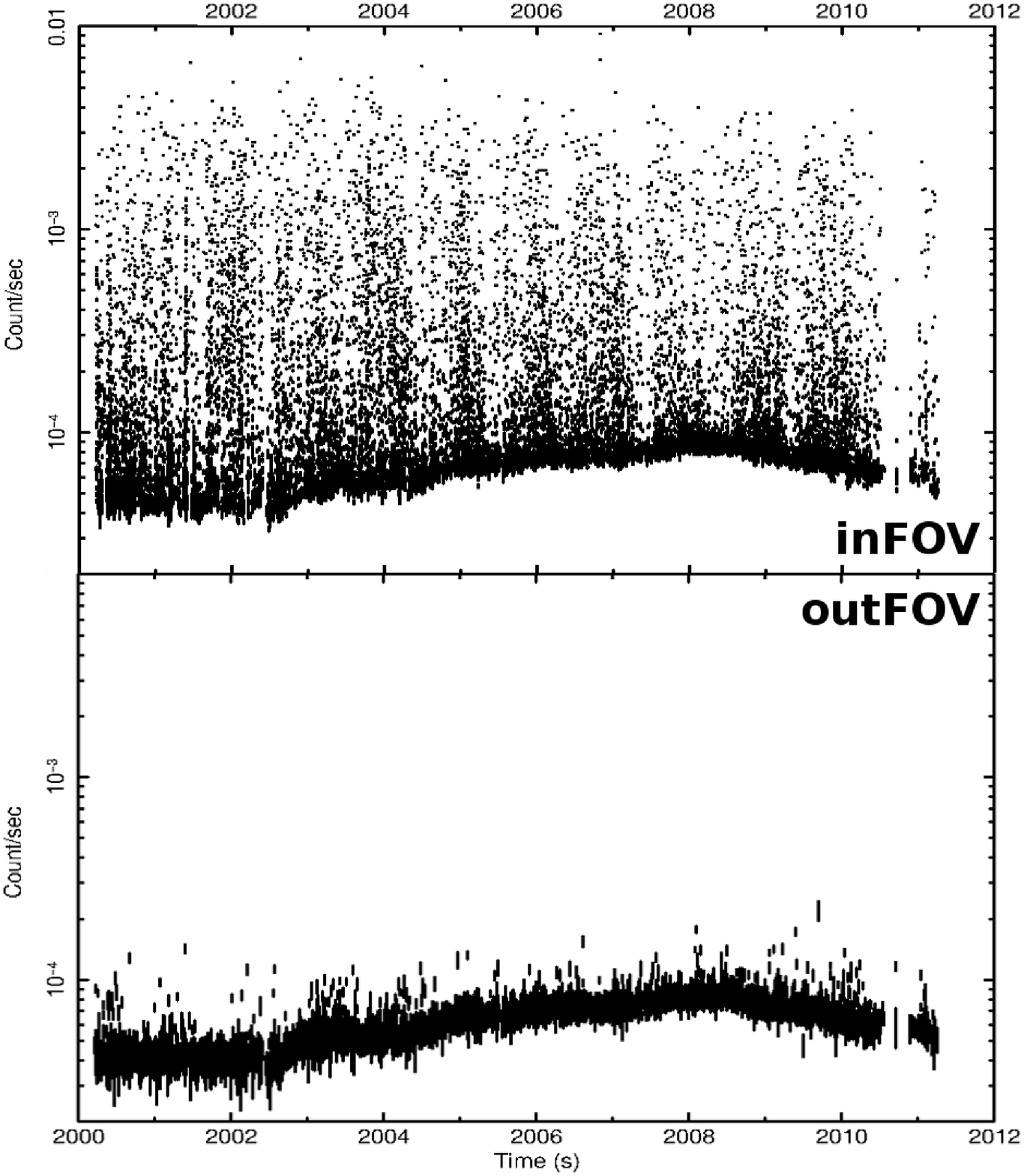}
\caption{Here, we show the light curves of inFOV (upper) and outFOV (lower) using our entire clean data set, 5000 s time bin. 1$\sigma$ errors are reported. Due to the area normalization and contaminants minimization of our method these curves are directly comparable; thus, the inFOV light curve shows the HEPI+SP background timing behaviour while the outFOV light curve shows the HEPI background timing behaviour.}
\label{fig-inof}
\end{figure*}

Using the ingredients described above, for each time bin of each raw light curve we compute the inFOV and outFOV clean rate and associated error as follows:

\begin{equation}
CR_I = R_I^{-1} dt^{-1} \sum_{j=1}^7 (\frac{N_j^I}{F_j})
\end{equation}
\begin{equation}
\sigma(CR_I) = R_I^{-1} dt^{-1} \sqrt{\sum_{j=1}^7 (\frac{N_j^I}{F_j^2})+\frac{3}{8}}
\end{equation}
\begin{equation}
CR_O = R_O^{-1} dt^{-1} \sum_{j=1}^7 (\frac{N_j^O}{F_j})
\end{equation}
\begin{equation}
\sigma(CR_O) = R_O^{-1} dt^{-1} \sqrt{\sum_{j=1}^7 (\frac{N_j^O}{F_j^2})+\frac{3}{8}}
\end{equation}
where $R_I$ is the inFOV rescale factor, $R_O$ is the outFOV rescale factor, $dt$ if the bin time, $N_j^I$($N_j^O$) the number of counts in the inFOV(outFOV) area of the j-th CCD, and $F_j$ the FRACEXP of the j-th CCD (see Section \ref{input} for more details).
Figure \ref{fig-lc} reports an example of the resulting background curve using standard SAS and our analysis tools.
The results for the entire data set are collected into an easy-readable single fits file (the $'$Main File$'$).
Figure \ref{fig-inof} reports the entire sample inFOV and outFOV light curves.

\section{Spectra computation}
\label{spectra}
\begin{figure*}
  \includegraphics[width=0.75\textwidth]{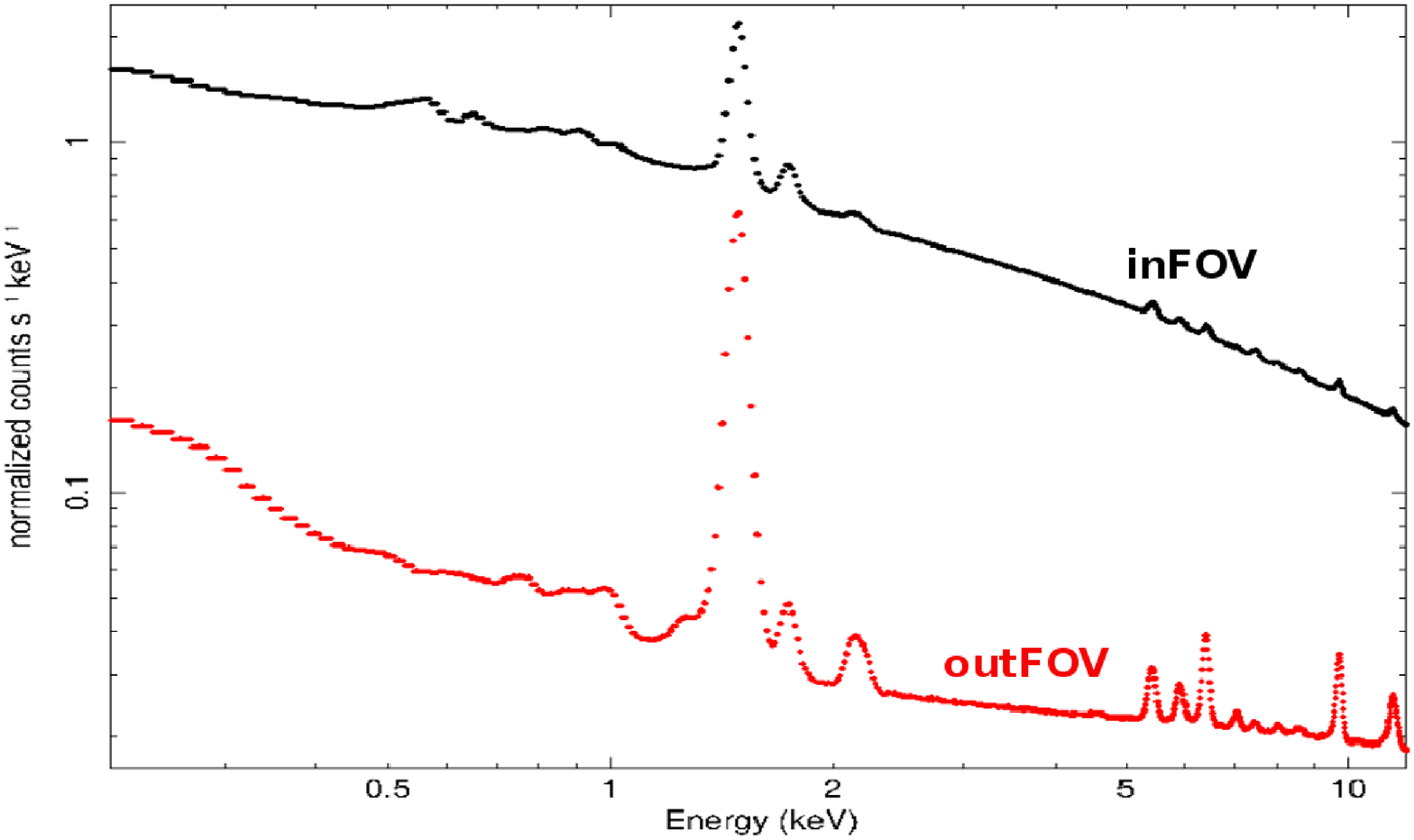}
\caption{Here, we show the spectra of inFOV (black) and outFOV (red) using our entire clean data set. We note that the two spectra are area- and exposure-corrected and therefore directly comparable.}
\label{fig-sp}
\end{figure*}

For each 500s time bin defined in Section \ref{lc}, we extract inFOV and outFOV spectra in the 0.2-12 keV energy range, using the same pattern, region and flags filters as for the light curve computation. We bin the spectra in order to obtain 15 channels/bin, for a total of 800 channels. For each raw of the light-curve Main File, we add a 800-elements array containing the grouped spectrum of that time bin, thus allowing any post-processing selection.
We wrote a tool that allows the user to make a selection of the 500-s spectra basing on a columns selection of the Main File (e.g. inFOV / outFOV ratio, optical filter, time, ...) and merges all the selected spectra into a single file for inFOV and one for outFOV. The BACKSCAL keyword in the spectrum (compliant with the standard OGIP format) takes into account the area rescale factor, so that they are rescaled to the same inFOV area. Similarly, the EXPOSURE keyword is correctly calculated. Figure \ref{fig-sp} reports, as an example, the inFOV and outFOV spectra of our entire clean dataset.

\section{Images computation}
\label{image}
\begin{figure*}9
  \includegraphics[width=0.75\textwidth]{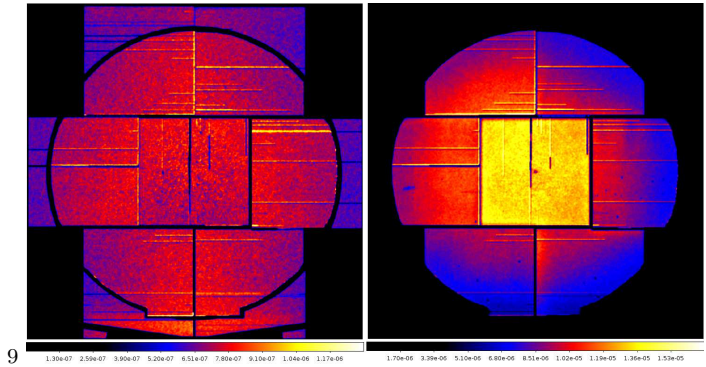}
\caption{An example of the images that we are able to produce. Here, we show the images of inFOV and outFOV regions for inFOV-outFOV values below 0.1 c s$^{-1}$ (left) and above 0.4 c s$^{-1}$ (right), thus roughly representing the high-energetic-particles induced background and the proton-flares-induced background we are analyzing in the AREMBES project. The different magnitudes of the two (flaring and quiescent) components do not allow for a direct comparison of the scales. }
\label{fig-im}
\end{figure*}

As a first step, we produce an exposure map for each observation, in detector coordinates and cheesed with the EXTraS background region (thus excluding contaminant celestial point-like sources). The image bin size is optimized to obtain both a good spatial resolution for the image and a reasonable size for the file.
Then, we extract images from each 500s light-curve time bin following the same filters as in Section \ref{lc} (also for the energy band). The results are stored in arrays as new columns of the Main File. In order to save disk space (and RAM), these results are compressed.
We wrote a tool that allows the user to make a selection of time bins based on columns of the Main File (e.g. inFOV / outFOV ratio, optical filter, time); it merges all the selected images into a single image for inFOV and one for outFOV. Moreover, this tool merges the corresponding cheesed exposure maps. The computed exposure-corrected images are therefore corrected for the excluded regions in the inFOV.
Figure \ref{fig-im} shows the MOS2 images of $'$quiescent$'$ and $'$flaring$'$ states.

While a detailed imaging analysis cannot be accommodated within the resources available to the AREMBES project, simple inspection of these images is sufficient to glean some rather interesting features. We list here some of them.\\
- There is a significant vignetting of the proton flares component (right panel), with a possible offset with respect to the center of the FOV.\\
- The proton flares component also presents a CCD-dependent spatial behaviour, with the central pixel brighter than the others (right panel).\\
- The quiescent background varies by about 10-20\% within each CCD increasing with distance from the read out node (left panel).\\
- Different CCDs appear to have different quiescent levels (left panel).\\

\section{Conclusions}

We reduced and analyzed the entire data set of {\it XMM-Newton} observations listed in the third {\it XMM-Newton} catalog, aimed at describing its particle-induced background. With $\sim$106 Ms of data, we reached an unprecedent level of accurancy with respect to the analysis in literature \cite{car07,kun08,lec08}.
Through event, pattern and energy selection we minimized contaminant effects such as CXB and detector noise.
Thanks to the EXTraS project products and newly-developed tools, we excluded celestial sources from the analysis down to an unprecedent level of accurancy.
This work allows for a complete characterization of soft-proton induced and high-energy-particles induced components of the {\it XMM-Newton} background, as well as for analysis of new possible focused and unfocused components.
For each 500-s time bin of the 106 Ms of data we extracted the corrected count rate, spectrum, image and exposure map and we stored informations into a fits file.
The results are area- and exposure-corrected, so that they are directly comparable.
Finally, we produced software tools that allow for the reconstruction of the products in every user-selected period.

\begin{acknowledgements}
The AHEAD project (grant agreement n. 654215) which is part of the EU-H2020 programm is acknowledged for partial support.
This work is part of the AREMBES WP1 activity funded by ESA through contract No. 4000116655/16/NL/BW.
Results presented here are based, in part, upon work funded through the European Union Seventh Framework Programme (FP7-SPACE-2013-1), under grant agreement n. 607452, “Exploring the X-ray Transient and variable Sky - EXTraS”.\\
Thanks to the reviewer for the useful comments. This resulted in a much improved work.
\end{acknowledgements}

\end{document}